# Evolving Neural Networks with Iterative Learning Scheme for Associative Memory *

**Shigetaka Fujita**
*Graduate School of Science and Technology*
*Kobe University*
*Rokkodai, Nada, Kobe 657, Japan*
e-mail: fujitan@hetsun1.phys.kobe-u.ac.jp

**Haruhiko Nishimura**
*Department of Information Science*
*Hyogo University of Education*
*Yashiro-cho, Hyogo 673-14, Japan*
e-mail: haru@life.hyogo-u.ac.jp

**Abstract**

A locally iterative learning (LIL) rule is adapted to a model of the associative memory based on the evolving recurrent-type neural networks composed of growing neurons. There exist extremely different scale parameters of time, the individual learning time and the generation in evolution. This model allows us definite investigation on the interaction between learning and evolution. And the reinforcement of the robustness against the noise is also achieved in the evolutional scheme.

# 1 Introduction

Recently evolutionary search procedures have been combined with the artificial neural networks (For example, see [1, 2] and references there in). In this type of models, information of the network structure and weights is encoded to the genotype, and the corresponding network is exposed to the selection by the environment as the phenotype. By taking the genotype/phenotype mapping indirect, the structure and weights can be the (unexpected) "emergent" properties of the network.

In our recent paper [3] we have proposed the biologically inspired model of the associative memory in recurrent-type artificial neural networks, based on the model of evolving neural networks (ENNs) [4, 5]. In this model the neurons are genetically placed in physical space and the connectivities are determined by the growth of their axons. That is, the mapping is set close to the neuronal developmental process. We found that the resulting physical network has highly asymmetric synaptic weights and dilute connections, but the ability for recalling the memorized patterns is sufficient in comparison with the fully connected Hopfield model.

In this paper we introduce a local iterative learning (LIL) rule for the phenotypic learning after the birth, instead of the simple Hebb's rule in our previous ENN model. The simple Hebb's rule has no definite algorithm to determine the learning time, due to the limitation of its synchronous stimuli. By using a LIL rule, we can evaluate the speed of learning to the desired stability goal. This enables us to study the effects of phenotypic learning stage on the genetic evolution and discuss the relationship between genotypic (instinctive) memories and phenotypic (acquired) ones, more explicitly.

# 2 Phenotypic Learning Stage

The learning stage during the life plays an important role for the ability of pattern retrieval, then the learning time (speed) should become the principal factor of the selection in evolution. Many LIL rules are known and have been investigated for the totally interconnected network until now [7]. Here, we take the perceptron learning rule [6, 8] which is a naive LIL rule and put it into our evolutionary scheme with partially and distortedly connected network.

Consider a neural network consisted of N neurons with synaptic weights $w_{ij}(w_{ii}=0,\, i,j=1,\ldots,N)$. Let $\xi_i^\mu(\mu=1,\ldots,p,\, i=1,\ldots,N)$ be a set of correlated $p$ patterns to be memorized and define the *stability coefficients* as

$$\gamma_i^\mu \equiv \xi_i^\mu \sum_{j=1}^{N} w_{ij}\xi_j^\mu. \tag{1}$$

The goal of the algorithm is to make the weights suit to the *stability condition*,

$$\gamma_i^\mu \geq \kappa \quad \text{for all } i,\mu, \tag{2}$$

iteratively. $\kappa$ ($> 0$) is the *minimum stability* which determines the depth of imbedding of the patterns. Let $w_{ij}(\tau)$ be the temporal weights in learning iteration, then $\gamma_i^\mu(\tau) = \xi_i^\mu \sum_{j=1}^N w_{ij}(\tau)\xi_j^\mu$, where $\tau$ is a counter of iteration which measures the convergence rate of the algorithm. Choose the first pattern ($\mu = 1$) and check weather $\gamma_i^\mu(\tau) \geq \kappa$ or not, for all $i$. If $\gamma_i^\mu(\tau) < \kappa$, then update the weights for all $j$ as

$$w_{ij}(\tau) \leftarrow w_{ij}(\tau) + \delta w_{ij}^\mu, \tag{3}$$

where

$$\delta w_{ij}^\mu = \frac{1}{N}(1 - \delta_{ij})\xi_i^\mu \xi_j^\mu. \tag{4}$$

Continue this stability check and weight update for all patterns until the last one ($\mu = p$). If the resulting $\gamma_i^\mu(\tau)$'s satisfy the stability condition, then the algorithm converges (learning is completed), else iterate the above process by setting $w_{ij}(\tau + 1) = w_{ij}(\tau)$.

When the stability condition of Eq. (2) is realized, it can be shown that each pattern is the fixed point of the network dynamics, whether the weights are asymmetric or not, and the network is partially connected or not. But, in the partially connected case, the robustness against the noise seems not to be assured as good as in the fully connected case. (This problem will be treated in section 4).

In the following section, we regard $\tau$ as the phenotypic *learning time* of a individual and try to include it in the fitness.

# 3 Model and Methods

This section outlines the model. For the details and our primitive motivation, see our previous paper [3].

Set a population containing 100 individuals. Each individual has a genotype encoded the information of creating their network structure. The genotype is divided into 49 blocks corresponding to 49 neurons. Each block contains the instructions for one neuron, namely, the "physical position" of the neuron ($x$ and $y$ coordinates in the "physical space". We take a square, normalized to $1 \times 1$, as the physical space), the directional angle $\alpha$ for the "axonal" growth, the axonal bifurcation angle $\theta$, the axonal bifurcation length $d$, the "presynaptic" weight $b$ (characterize the firing intensity and neuron's type, excitatory or inhibitory) and the "postsynaptic" weight $a$ (amplify or damp the incoming signal). All values are randomly generated at the 0-th generation with the ranges $0 \leq x, y \leq 1$, $0 \leq \alpha \leq 2\pi$, $0 \leq \theta \leq \pi/3$, $0 \leq d \leq 0.15$, $-2 \leq b \leq 2$ and $0 \leq a \leq 2$.

From each neuron in the physical space, an axon grows and bifurcates as shown in Fig. 1 (in the simulation, five times) and connects with other neurons. The connections of neurons are determined by whether the axon reaches the neighbourhood $r$ (fixed to $r = 0.05$) of other neurons or not. So the connections of neurons are generally partial. We introduce the connectivity matrix $c_{ij}$ $(i, j = 1, \ldots, N;\ c_{ii} = 0)$ as $c_{ij} = 1$ if the axon from the $j$-th neuron reaches the $i$-th neuron, and otherwise $c_{ij} = 0$. Then the *bare* synaptic weight $w^0_{ij}$ at birth, before learning, is given by $w^0_{ij} = a_i b_j$ if $c_{ij} = 1$, and $w^0_{ij} = 0$ if $c_{ij} = 0$. The bare synaptic weights are generally asymmetric ($w^0_{ij} \neq w^0_{ji}$).

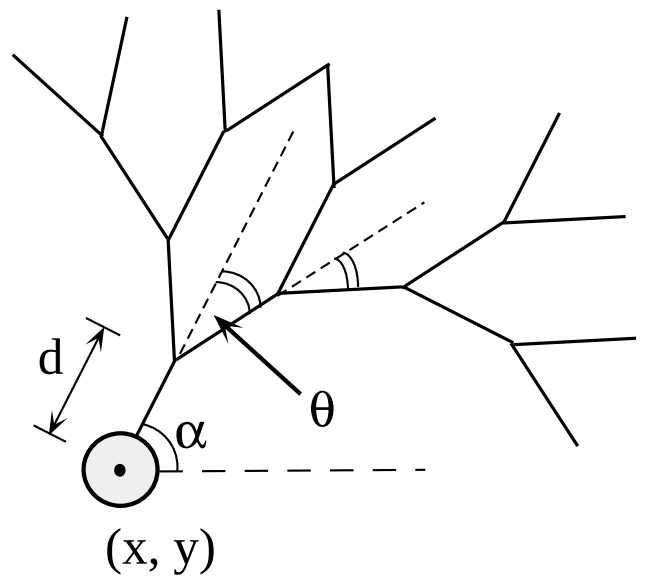

Figure 1: The Growing and Bifurcating Process of an Axon.

After the bare weights are established, only the weights with $c_{ij} = 1$ are *dressed* by the phenotypic learning in section 2, starting from $w_{ij}(0) = w^0_{ij}$ for $\kappa = 1$ (*learning task*). We use the 7×7 non-orthogonal four ($p = 4$) patterns as a set of target patterns (Fig. 2). When the stability condition of

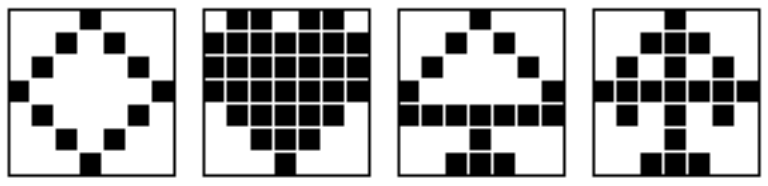

Figure 2: An example of a set of target patterns.

Eq. (2) is satisfied by $\tau$-time iterations (if not, up to $\tau_{max} = 100$ times), the dressed weights are obtained and its *learnability* (learning speed) is measured by the form of $(\tau_{max} - \tau)/\tau_{max}$. And next the *retrieval task* is imposed to the network having the dressed weights $w_{ij}$. Concretely, one of the four patterns with random noise is given to the individual network as the initial states of neurons. And the states of neurons are updated synchronously according to the equation $s_i(t+1) = f(\sum_{j=1}^{N} w_{ij} s_j(t))$, where $s_i(t)(-1 \leq s_i \leq 1)$ is the state of the $i$-th neuron and $t$ is the time step at retrieving. The transfer function $f(x)$ is taken to be $f(x) = tanh(x/2\epsilon)$. The steepness parameter $\epsilon$ is set to 0.015. During the update proceeds, we trace the overlap of the network state with the target pattern: $m^\mu(t) = \frac{1}{N}|\sum_{i=1}^{N} \xi_i^\mu s_i(t)|$ at the step $t$=5,10 and 15. This retrieving trial is applied to all patterns. In order to estimate the

*retrievability* we use

$$m = \frac{1}{4}\sum_{\mu=1}^{4}(m^{\mu}(5) + m^{\mu}(10) + m^{\mu}(15))/3, \quad (5)$$

which is the same as in our previous paper. The above retrieval task is repeated four times successively by changing the noise to get the balanced robustness. So, $m$ of Eq. (5) averages to $\overline{m}$. Finally we define the fitness $F$ as

$$F = 100 \times \frac{\tau_{max} - \tau}{\tau_{max}} + 100 \times \overline{m} \quad (6)$$

which reflects both the learnability (learning speed) and the retrievability.

At the end of one generation, the best 20 of individuals which have got higher fitness values are selected, and 5 copies of each of their genotypes are inherited to their offspring with random mutation (mutation rate is 0.004. we do not use crossover for simplicity). The above process continues for appropriate generations.

# 4 Results of Simulations

Bold solid lines in Fig. 3 show the changes of the fitness $F$ of the best individual until the 1000th generation in the cases of 0.05 noise (Fig. 3(a)) and no (0.00) noise (Fig. 3(b)) for the retrieval task, respectively. The thin solid line represents the fitness of the learnability (the former part of F) and the dashed line represents the fitness of the retrievability (the latter part of F). Comparing with the "pure genetic" result which shows the ability of genetic algorithm without learning in our previous paper [3], we can observe that the evolution of the retrievability is much improved owing to the existence of phenotypic learning stage. Furthermore, we find that after the retrievability reaches its near maximal value (100), the learnability (learning speed) keeps rising. That is, the function of dressed weights are replaced gradually with that of bare weights in subsequent generations. From the aspect of the Baldwin effect [9, 10], in which abilities that initially require learning are eventually replaced by the evolution of genetically determined systems without learning, we can accept this tendency naturally.

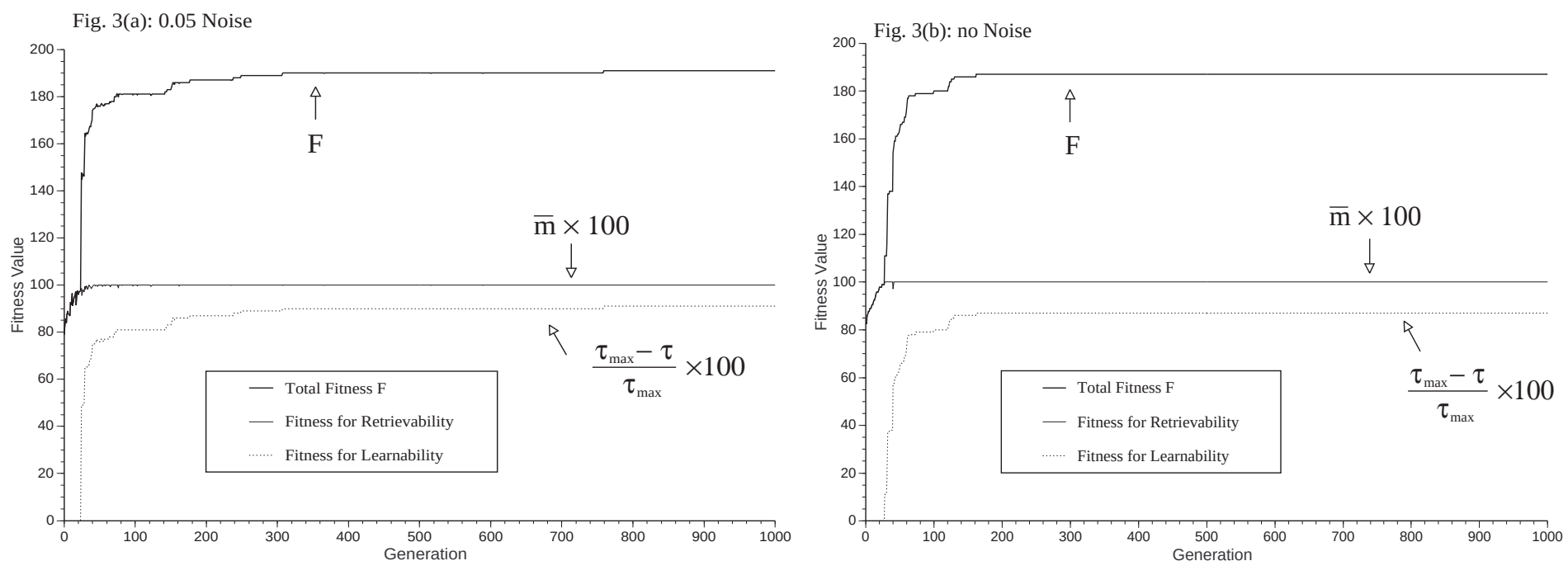

Figure 3: Evolution of fitness of the best individual in population. (a) 0.05 noise. (b) no (0.00) noise.

Fig. 4 shows the transition of the connection rate (= 1 if totally connected) of the best individual until the 1000th generation in the cases of 0.05 noise and no noise. In Fig. 3 there is not a great difference in fitness values between the two cases, but in the connection rate there exists a good difference. The higher task pressure with noise makes the evolution of individuals increase the connections.

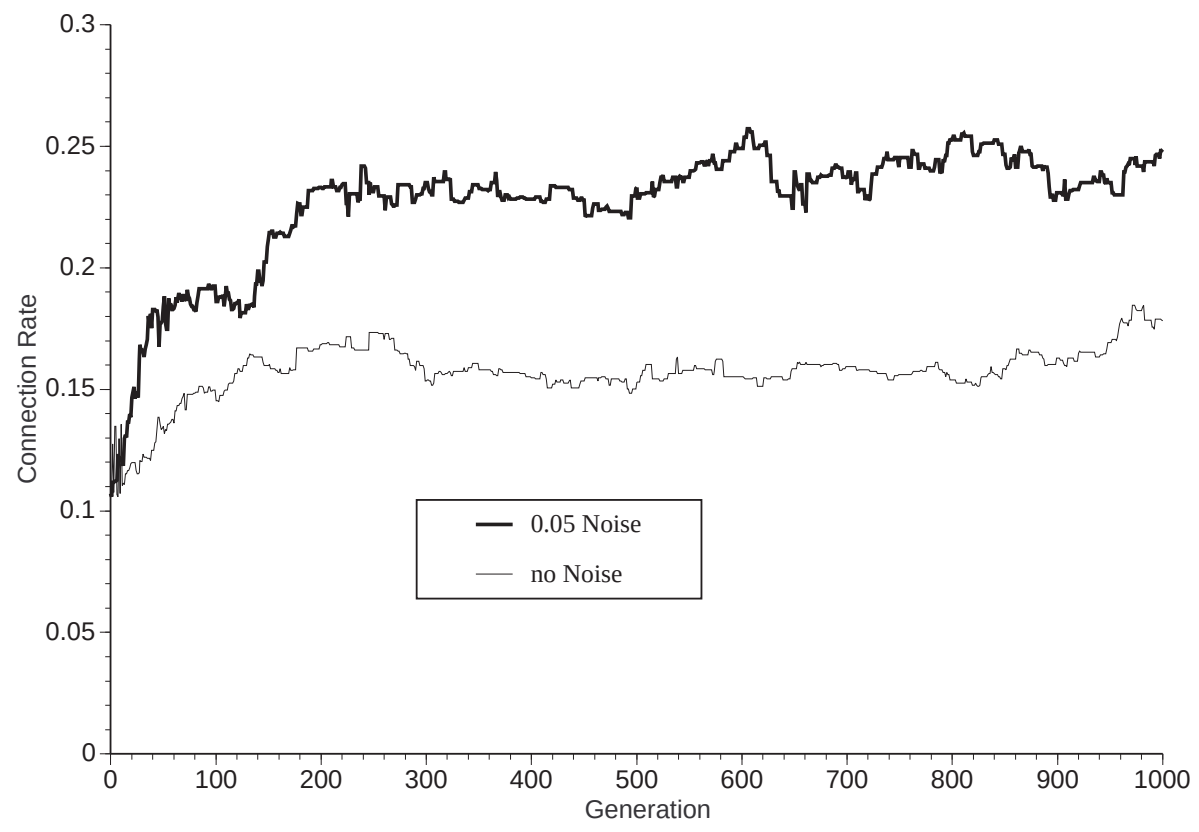

Figure 4: Transition of connection rate of the best individual in the cases of 0.05 noise and no noise.

Next we check the network robustness of retrieval against the noise level, imposing 100 retrieval tests in each level to the best individual at the 1000th generation. From Fig. 5 it is found that the resulting individual selected under the retrieval task with 0.05 noise (bold solid line) has got better robustness (beyond 0.05 level). In spite of its dilute connection, the performance is comparable to the conventional non-evolutionary model which is totally connected learned by the perceptron rule. On the other hand, the individual selected under the retrieval task without noise (thin solid line) can not bear the increasing noise level. This result is close to that of the only learning task case. Then to see the pure learnability (learning speed), we were practically obliged to add an auxiliary term to the fitness, because our genetic space for search is very large and difficult to find the expected individual by random search without the retrieval requirement. In fact, we have tested the random search by generating randomly 100000 genotypes and imposing them the learning with $\tau_{max} = 1000$, but we could not find the expected individual which satisfies $\gamma_i^\mu \geq 1$ for all $i$ and $\mu$. So we have taken $n_\gamma$, the number of the stability coefficients with $\gamma_i^\mu \geq 1$ as the auxiliary term and used $\frac{n_\gamma}{4\times 49} + \frac{\tau_{max}-\tau}{\tau_{max}}$ as the fitness.

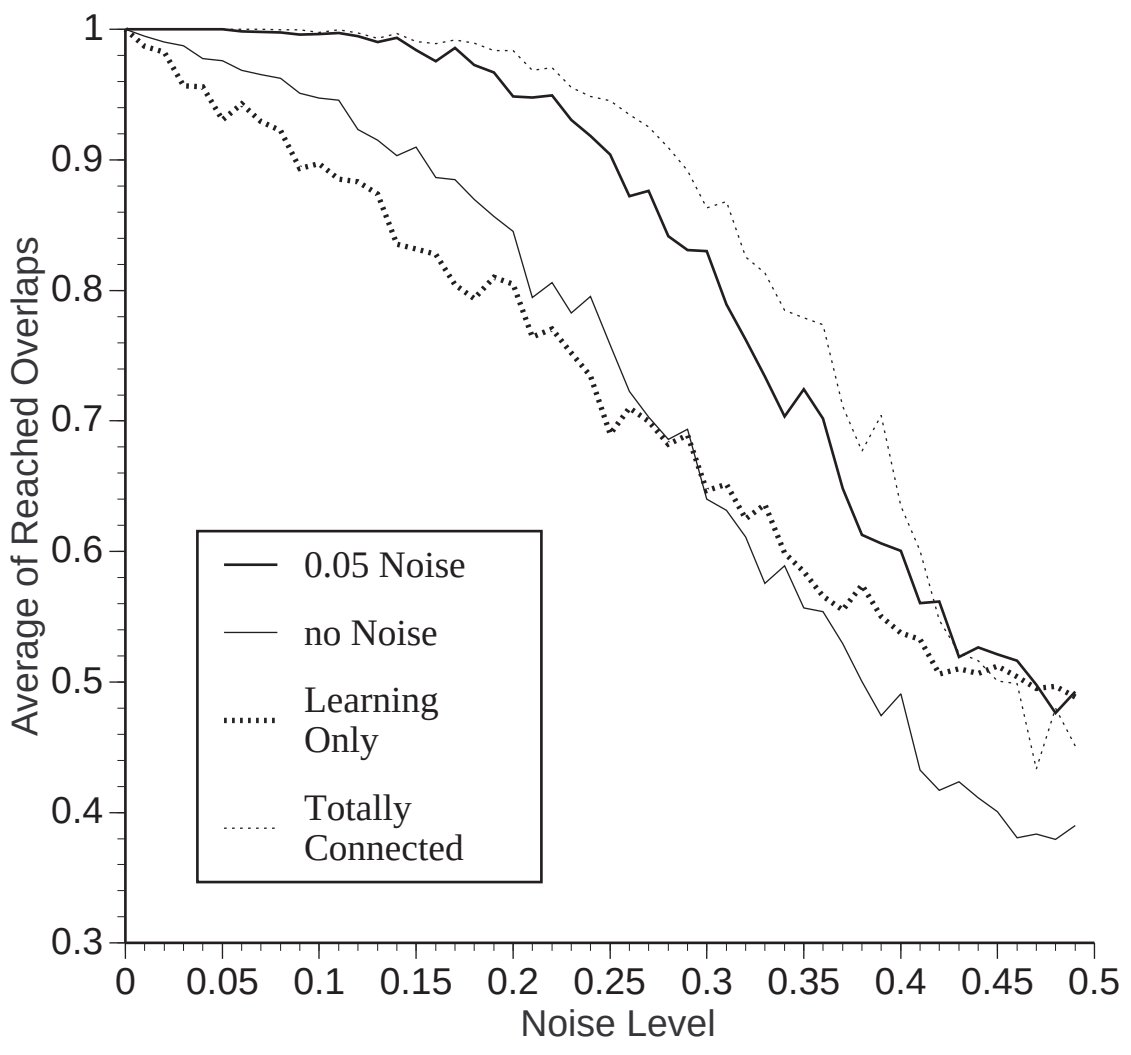

Figure 5: Robustness of the best individual against various noise.

Changes of the network structure of the best individual is shown in Fig. 6 from the 0th generation(G0) to the 1000th(G1000) in the case of 0.05 noise task. At G1000 the value of symmetry parameter ($\eta = \Sigma_{i,j} w_{ij} w_{ji} / \Sigma_{i,j} w_{ij}^2$) is given by $\eta = 0.024$. There seems to be some characteristics of self organization as a physical object in the space, but further analyses are needed for definite discussion.

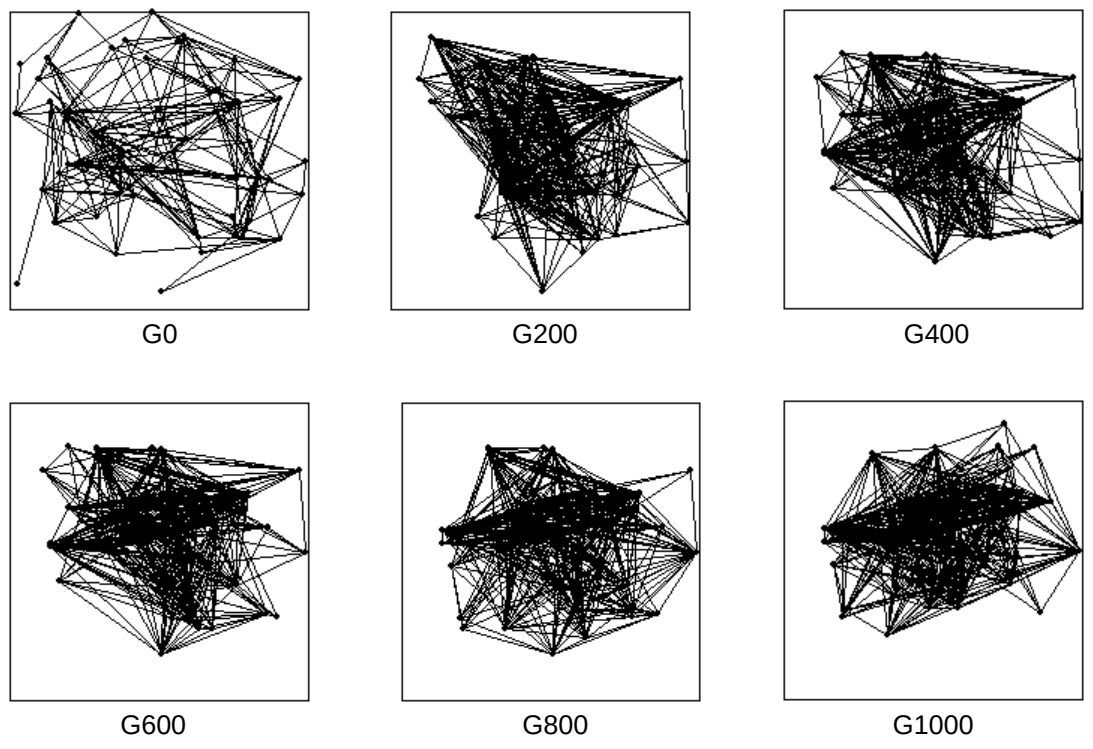

Figure 6: Changes of the network structure of the best individual from G0 to G1000 in the case of 0.05 noise.

# 5 Conclusion

We have presented the results of the simulations of associative memory with the LIL rule in the recurrent-type neural networks composed of growing neurons.

By combining the LIL rule as phenotypic learning stage to the ENN, the model becomes to have extremely different scale parameters of time. One is the individual learning time $\tau$ and the other is the generation of evolution. We found that the robustness against the noise is obtained through the interaction between the learnability (learning speed) and retrievability. The tendency that phenotype dependent (RAM-like) memories change into genotype dependent (ROM-like) memories can be interpreted as the Baldwin effect.

To bring the LIL rule into the model by hand is not natural and we think the learning rule itself should emerge in evolution (learning to learn). This problem is left to further investigation.